\documentclass{mem}
\usepackage{natbib}\usepackage{txfonts}\usepackage{balance}
\usepackage{graphicx}
\idline{75}{1}
\begin{document}
\def\teff{$T\rm_{eff }$}
\def\kms{$\mathrm {km s}^{-1}$}

\newcommand{\mincir}{\raise
-3.truept\hbox{\rlap{\hbox{$\sim$}}\raise4.truept\hbox{$<$}\ }}
\newcommand{\magcir}{\raise
-3.truept\hbox{\rlap{\hbox{$\sim$}}\raise4.truept\hbox{$>$}\ }}

\title{The Environment of Sy1, Sy2 \& Bright 
IRAS Galaxies: The AGN/Starburst connection.}

   \subtitle{}

\author{E.Koulouridis\inst{1,3}, V.Chavushyan\inst{2}, 
M.Plionis\inst{1,2}, D.Dultzin\inst{4}, 
Y.Krongold\inst{4}, C.Goudis\inst{1,3}, E.Chatzichristou\inst{1}
          }

\institute{Institute of Astronomy \& Astrophysics, National Observatory of
Athens, I.Metaxa \& B.Pavlou, P.Penteli 152 36, Athens, Greece
\and
Instituto Nacional de Astrofisica, Optica y Electronica (INAOE)
 Apartado Postal 51 y 216, 72000, Puebla, Pue., Mexico
\and
Physics Department, Univ. of Patras, Panepistimioupolis Patron, 26500, Patras, 
Greece
\and
Instituto de Astronom\'ia, Univesidad Nacional
Aut\'onoma de M\'exico, Apartado Postal 70-264, M\'exico, D. F.
04510, M\'exico
}

\authorrunning{Koulouridis}

\titlerunning{The Environment of Sy1, Sy2 \& Bright 
IRAS Galaxies}

\abstract{
We present a 3-dimensional study of the local  ($\leq 100 \; h^{-1}$
kpc) environment of Sy1, Sy2 and Bright IRAS Galaxies.
For this purpose we use three galaxy samples (Sy1, Sy2, BIRG) located at
high galactic latitudes as well as three control sample of non-active
galaxies having the same morphological, redshift and diameter size
distributions as the corresponding Seyfert or BIRG sample.
Using the CfA2 and SSRS galaxy catalogues as well as our own 
spectroscopic observations, we find that the fraction of BIRGs 
with a close neighbor is significantly 
higher than that of their control sample. We also find that Sy2 galaxies 
demonstrate the same behaviour with BIRG galaxies but not with Sy1s which do 
not show any excess of companions with respect to their control sample 
galaxies. 
An additional analysis of the relation between FIR colors and activity type
of the BIRG's shows a significant difference between 
the colors of strongly-interacting and non-interacting starbursts 
and a resemblance between the colors of non-interacting starbursts and Sy2s.
Our results support an evolutionary scenario leading from Starbursting to 
a Sy2 and finally to an unobscured Sy1 galaxy, 
where close interactions play the role of the triggering mechanism.

}
\maketitle{}

\section{Introduction}

Despite the numerous studies conducted over the last decade 
investigating the 
relation among interacting galaxies, starbursting and 
nuclear activity (eg. Hernandez-Toledo et al. 2001;  Ho 2005, Gonz\'alez et al.
 2008), 
the correlation between these physical processes remain
uncertain. However, there is 
evidence that AGN galaxies host a post-starburst stellar 
population (eg. Gonz\'alez Delgado 2001, Kaviraj 2008, Muller Sanchez 2008)
while Kauffmann et al. (2003) showed that the
fraction of post-starburst stars increases with AGN emission. 
Proving such a relation
would simultaneously solve the problem of the AGN triggering
mechanism. Interactions would be the main cause of such activities, being
starbursting and/or the feeding of a central black hole. 
Indeed some studies seem to conclude that there is an evolutionary 
sequence from starburst to type 2 and then to type 1 AGN galaxies 
(e.g. Oliva et al. 1999, Storchi-Bergmann et al. 2001, Krongold et al. 2002, 
Chatzichristou 2002). 
In addition, Kim, Ho and Im (2005)
using the [OII] emission line as a SFR indicator, reach the conclusion 
that type 2 are the precursors of type 1 quasars supporting the previous 
claims. Finally, a study by Ya-Wen Tang et al. (2008) tracing 
interactions by HI imaging, showed that 94\% of their 
Seyfert sample exhibit HI disturbances in contrast with their inactive 
counterpart (15\%). 
The above results raise doubts about the simplest version 
of the unification scheme of AGNs.

\section{Observations \& Samples}

The Bright IRAS sample consists of 87 objects with 
redshifts between 0.008 and 0.018 and was compiled 
from the BIRG survey by Soifer et al. (1989) for the 
northern hemisphere and by Sanders et al. (1995) for 
the southern. More details about the sample 
selection are given in Koulouridis et al. (2006b). 
We also use the control sample, compiled by Krongold et al. (2002) in
such a way as to reproduce the main characteristics, other than the
infrared emission, of the Bright IRAS sample. 

The samples of the two type of Seyfert galaxies were compiled from the
catalog of Lipovetsky, Neizvestny \& Neizvestnaya (1988). They consist of 
72 Sy1 galaxies with redshifts between 0.007 and 0.036 and 69 Sy2 galaxies 
with redshifts 
between 0.004 and 0.020. The samples are volume limited as indicated
from the $V/V_{max}$ test and complete to a 
level of 92\%. The sample selection details are described in Dultzin-Hacyan 
et al. 1999 (hereafter DH99) and Koulouridis et al. (2006a). 
We also use the two control samples, compiled by DH99 in
such a way as to reproduce the main characteristics, other than the
nuclear activity, of the AGN samples. 
The use of control samples is very important in order to be able to attribute
the nuclear/starbursting activity to possible
environmental effects and not to sample biases or possible 
host galaxy differences.

The redshift distribution of Seyfert galaxies  peaks at a slightly
higher redshift than that of BIRG galaxies 
imposing the Seyfert magnitudes to be relatively closer to the 
limit of the CfA2 and SSRS catalogues. Therefore 
we may be missing neighbors of a similar magnitude difference between
them and the central BIRG or Seyfert galaxy.
Note however that such a difference will not influence the comparison 
between the BIRG or Seyfert galaxies and their respective control sample 
galaxies (since both have the same redshift distribution).

In order to cover a larger magnitude difference between Seyferts/BIRGs and
their neighbors than that imposed by the CFA2/SSRS magnitude
limit ($m_B\sim 15.5$) we obtained our own spectroscopic observations
of fainter neighbors around three subsamples 
consisting of 22 Sy1, 22 Sy2 and 24 BIRG galaxies 
(selected randomly from our samples). Around each galaxy we have 
obtained spectra
of all galaxies within a projected radius of 100 $h^{-1}$ kpc 
and a magnitude limit of $m_B\sim 18.5$. 

Optical spectroscopy was carried out using the Faint
Object Spectrograph and Camera (LFOSC)
mounted on the 2.1m Guillermo Haro telescope in Cananea,  
operated by the National Institute of Astrophysics, Optics and 
Electronics (INAOE) of Mexico. A setup covering the spectral range 
$4200-9000$\AA\ with a dispersion of 8.2\,\AA/pix was adopted. 
The effective instrumental spectral resolution was about 15\,\AA. 
The data reduction was done using the IRAF packages. 

\section{Analysis and Results}
We identify the nearest neighbor of each Seyfert, Bright IRAS
and control galaxy in our samples with the aim of estimating the fraction
of galaxies that have a close neighbor. To define the
neighborhood search we use two parameters, 
the projected linear distance ($D$) and the radial velocity        
separation ($\delta u$) between the central galaxy and the neighboring
galaxies found in the CfA2 and SSRS
catalogues or in our own spectroscopic observations.
We search for neighbors with
$\delta u \leq$ 600 km/s, which is roughly the
mean galaxy pairwise velocity of the CfA2 and SSRS galaxies or about 
twice the mean pairwise galaxy velocity when clusters of galaxies are
excluded. 
We then define the fraction of active and non-active galaxies that
have their nearest neighbor, within the selected $\delta u$
separation, as a function of increasing $D$.

\subsection{Neighbors analysis}
In Figure 1 we plot the fraction of Bright IRAS, Seyfert and control galaxies 
that have a close companion, as a function of the projected distance ($D$) 
of the first companion. We present results for $\delta u\le 200$
km/s (left panel) and $\delta u\le 600$ km/s (right panel). 

\begin{figure}[h]
\resizebox{\hsize}{!}{\includegraphics[clip=true]{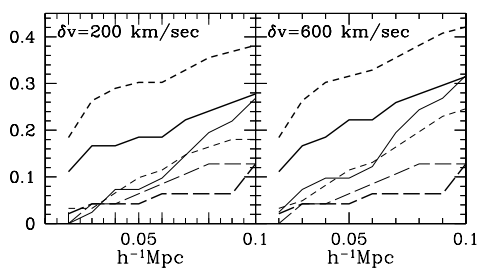}}
\caption{
\footnotesize
Fraction of BIRGs (thick short dashed line), Sy2s (thick line), 
Sy1s (thick long dashed line) and their respective control 
sample galaxies (thin lines) which have their nearest neighbor 
within the indicated redshift separation,
as a function of projected distance. Uncertainties are 
$\sim \pm 0.05$ in 
fraction.}
\end{figure}

It is evident that the Sy1 galaxies and their control sample
show a consistent fraction of objects having a close neighbor (within
the errors). On the other hand, there is a significantly higher fraction of Sy2
galaxies having a near neighbor, especially within $D\leq 75$ h$^{-1}$ kpc,
with respect to both their control sample and the Sy1 galaxies. 
This confirms previous results based on a two dimesional analysis (DH99).
Adding the BIRG sample, which includes mostly starburst and 
Sy2 galaxies, we can clearly see that an even 
higher fraction of BIRGs tend to have a close companion  
within $D\mincir 75$ h$^{-1}$ kpc.

In order to investigate whether fainter neighbors, than those found in
the relatively shallow CFA2 and SSRS catalogues, exist around our
BIRGs, we have analyse our
spectroscopic survey of all neighbors with $m_B\mincir 18.5$
and $D\le 75 \; h^{-1}$ kpc for 
a subsample of 22 Sy1, 22 Sy2 and 24 BIRG galaxies.
We find that the percentage of 
both Sy1 and Sy2 galaxies that have a close neighbor increases 
correspondingly by about 100\% 
when we descent from $m_B\mincir 15.5$ to $m_B\mincir 18.5$, 
while the percentage of BIRGs rises only by $\sim$45\% reaching the
equivalent Sy2 levels. 

\subsection{FIR color analysis} 

In these section we investigate whether there is relation 
between the strength of the interaction of BIRGs with their
 closest neighbor and their FIR characteristics. The strength
 of any interaction could be parametrized as a function of the distance 
between the BIRG and its first neighbor. 
We divided the interactions in our sample 
into three categories based on the proximity of the first 
neighbor. We consider strong interactions when 
$D\leq 30 \; h^{-1} \; kpc$, weak interactions when $30\leq D\leq 100
\; h^{-1} \; kpc$ and no interaction when $D> 100 \; h^{-1} \; kpc$.  

In Figure 2 we present the color - color diagram of $\alpha(60,25)$ 
versus $\alpha(25,12)$, where $\alpha(\lambda_1, \lambda_2)$ is 
the spectral index defined as 
$\alpha(\lambda_1,
\lambda_2)=log(S_{\lambda_1}/S_{\lambda_2})/(\lambda_2/\lambda_1)$ 
with $S_{\lambda_1}$ the flux in janskys at wavelength $\lambda_1$. 
We can clearly see the differences between the FIR characteristics 
among different types of galaxies and different interaction strengths.

It is evident that the FIR characteristics of starburst galaxies in 
our BIRG sample differ significantly depending on the strength of the
 interaction. The majority of
 highly interacting starburst have $\alpha(60,25)$ spectral indices greater 
than -2,
 while all, except one, non-interacting starbursts have less. We also find
that normal galaxies and Liners are at the lower end of this sequence. 
However, Sy2 galaxies, interacting or not, 
seem to lay in the same area
with non-interacting starburst galaxies, 
delineated by two dashed lines.

The FIR color analysis of our sample strengthens our previous results. It 
clearly shows that the starburst activity is higher when
interactions are stronger and ceases when the interacting neighboring 
galaxy moves away. While the starburst activity weakens (if we link position on
the plot with time) Sy2 nuclei appear, giving further evidence on the causal
bridging between these objects.

\section{Discussion \& Conclusions}

\begin{figure}[t]
\resizebox{\hsize}{!}{\includegraphics[clip=true]{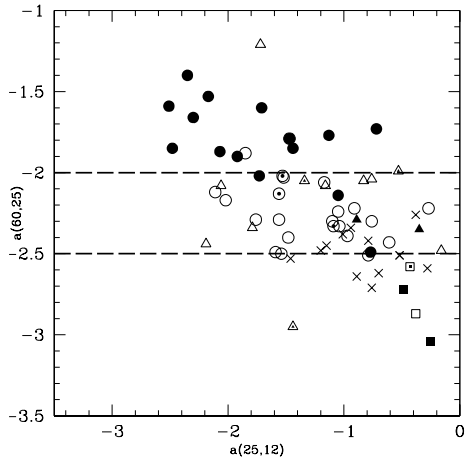}}
\caption{
\footnotesize
FIR color-color diagram: $\alpha(60,25)$ versus
  $\alpha(25,12)$ : Starbursts are
  represented by circles, Sy2s by triangles, Liners by squares and normal
  galaxies by crosses. Highly interacting galaxies are represented 
by filled shapes, weakly interacting by open dotted shapes 
and non-interacting by open shapes.}
\end{figure}

Our results can be accommodated in a simple 
evolutionary scenario, starting with an interaction, and ending in a Sy1.
First, close interactions would drive molecular clouds 
towards the central area, creating a circumnuclear starburst. Then, material
could fall even further into the innermost regions of the galaxy, feeding the
black hole, and giving birth to an AGN which at first cannot be observed due to
obscuration. At this stage only a starburst would be observed. As starburst
activity relaxes and obscuration decreases, a Sy2 nucleus would be revealed
(still obscured by the molecular clouds from all viewing angles). As a
final stage, a Sy1 would appear. In this case, the molecular clouds, 
initially in a spheroidal distribution, 
could flatten and form a ``torus'' (as in the
unification scheme for Seyferts). As more material is accreted, it is possible
that the AGN strengthens driving away most 
of the obscuring clouds, and leaving a ``naked'' Sy1 nucleus. 

In order to better understand the role of interactions in driving starburst
and nuclear activity, we are in
the process of completing a study of AGN and starburst manifestations in
the nearest neighbors of the active galaxies in our samples.
Preliminary results show that more than 70\% of the close companions
of both type of Seyfert galaxies show some kind of activity. Most 
companions are 
HII galaxies and only a few are Seyferts. Furthermore, the fraction O[III]/Hb 
is significally higher in all Sy2 companions indicating a higher ionization.
The correlation between high ionization and galaxy interactions is relatively
well established for HII galaxies (eg. Woods, Geller \& Barton 2007), 
although Alonso et al. (2007) presents such a connection also 
for AGNs.

\begin{acknowledgements}
EK thanks the IUNAM and INAOE for its warm hospitality. 
MP acknowledges funding by the Mexican Government research grant
No. CONACyT 49878-F, VC by the CONACyT research grant 54480 and DD    
support from grant IN100703 from DGAPA, PAPIIT, UNAM.
\end{acknowledgements}

\end{document}